\documentclass{elsarticle}
\usepackage{graphicx}%
\usepackage{multirow}%
\usepackage{amsmath,amssymb,amsfonts}%
\usepackage{amsthm}%
\usepackage{mathrsfs}%
\usepackage[title]{appendix}%
\usepackage{xcolor}%
\usepackage{textcomp}%
\usepackage{manyfoot}%
\usepackage{booktabs}%
\usepackage{algorithm}%
\usepackage{algorithmicx}%
\usepackage{algpseudocode}%
\usepackage{listings}%
\usepackage{verbatim}
\usepackage{enumitem}
 \usepackage{comment}

\journal{}
\begin{document}

\begin{frontmatter}

\title{Non-equilibrium phase transition and cultural drift in the continuous-trait Axelrod model }

\author[UFPE]{Paulo R.  A. Campos}

\author[GMU]{Sandro M. Reia}

\author[mymainaddress]{Jos\'e F.  Fontanari}

\address[UFPE]{Departamento de F\'{\i}sica, Centro de Ci\^encias Exatas e da Natureza,  Universidade Federal de Pernambuco,  50740-560 Recife,  Pernambuco,  Brazil}

\address[GMU]{Geography and Geoinformation Science, College of Science, George Mason University, Fairfax, VA, USA}

\address[mymainaddress]{Instituto de F\'{\i}sica de S\~ao Carlos,  Universidade de S\~ao Paulo,  13566-590 S\~ao Carlos,  S\~ao Paulo,  Brazil}

\begin{abstract}
The standard Axelrod model of cultural dissemination, based on discrete cultural traits, exhibits a non-equilibrium phase transition but is inherently limited by its inability to continuously probe the critical behavior. We address this limitation by introducing a generalized Axelrod model utilizing continuous cultural traits confined to the interval $[0,1]$,  and a similarity threshold,  $d$,  that serves as a continuous control parameter representing cultural tolerance. This framework allows for a robust analysis of the model's critical properties and its dynamics under cultural drift (copying noise). For the perfect copying scenario, we precisely locate the critical threshold   $d_c$, which separates the disordered (fragmented) and ordered (polarized) phases. Through Finite-Size Scaling, we  find that the mean domain density vanishes continuously at $d_c$ with the  exponent   $\beta = 1/3$. Simultaneously, the largest domain fraction displays a surprising discontinuous jump at $d_c$.   
We find that the finite size effects in the critical region are governed by the exponent $\nu=2$ for both the continuous and discontinuous transitions.
Under  imperfect copying,  persistent noise introduces a powerful selective pressure on the trait space, leading to the emergence of two symmetry-related attractors at the trait values $d$ and $1-d$.  However, these noise-induced attractors prove fragile in the thermodynamic limit, becoming unstable at large lattice sizes, which directly accounts for the observed failure of the dynamics to freeze under sustained cultural drift. This suggests that in large, continuously evolving societies, true cultural convergence is highly unlikely,  leading instead to sustained fragmentation and nonstationary dynamics where cultural domains never fully stabilize.
\end{abstract}




\end{frontmatter}

\section{Introduction}\label{sec1}

The study of cultural change and the maintenance of diversity in societies has been approached through two major modeling traditions: cultural transmission models \cite{Boyd_1985} and agent-based models of social influence \cite{Epstein_1996, Goldstone_2005, Castellano_2009}. Classical cultural transmission models, rooted in population genetics, primarily focus on the aggregate dynamics of cultural traits over generations. These are driven by mechanisms like vertical transmission (parent to child) and various social learning biases, such as oblique or horizontal transmission, where individuals learn from non-parental role models \cite{Bisin_2001, Buechel_2014, Fogarty_2024}. While these frameworks successfully demonstrate how cultural traits can be sustained, the process of interaction and influence is often implicitly assumed or defined globally. In contrast, the agent-based model for the dissemination of culture proposed by political scientist Robert Axelrod fundamentally shifts the focus to local interaction rules on a social network \cite{Axelrod_1997}. The core innovation of the Axelrod model is the explicit embedding of the homophily principle---the idea that similarity breeds influence. In this framework, influence (the social learning event) is directly contingent upon existing similarity: the probability of two agents interacting is determined by their cultural overlap. This local, homophily-driven convergence mechanism is precisely what allows the Axelrod model to generate the intriguing macro-phenomenon of global polarization, making it a cornerstone for understanding how fragmented cultures persist despite strong local tendencies toward assimilation.

The Axelrod model's simplicity and non-trivial behavior immediately attracted the attention of the statistical physics community. This adoption transformed the model into a standard benchmark in the field of non-equilibrium statistical mechanics, owing to the sharp and non-trivial phase transition it exhibits between the ordered (polarized) and disordered (fragmented) states \cite{Castellano_2000, Vilone_2002, Klemm_2003, Klemm_2003b}. Beyond its theoretical appeal, the model's core principle of homophily provides a powerful framework for describing a variety of real-world phenomena where similarity drives interaction and subsequent change. These applications include the diffusion of innovations \cite{Tilles_2015, Pineda_2023}, the effects of mass media influence \cite{Shibanai_2001, NJP_2010, Peres_2011}, and collective problem solving \cite{Kennedy_1998, Fontanari_2010}. The model's success lies in its ability to generate complex, macroscopic patterns from a simple, microscopic rule, making it an invaluable tool for physicists and social scientists alike.

Despite the success of the original Axelrod model, the use of discrete cultural traits presents both conceptual and technical limitations. Conceptually, many real-world cultural attributes, such as opinions, beliefs, and economic investments, exist on a continuous spectrum rather than being limited to a small number of nominal categories. The transition to a continuous trait space, which we adopt here, has already been embraced by advanced cultural evolution models (e.g., in the intergenerational framework \cite{Bisin_2001, Buechel_2014}). Technically, the discrete nature of the traits in the original model, which is characterized by the number of nominal categories $q$, makes a precise analysis of the phase transition difficult. While variants like the Poisson model address this issue \cite{Castellano_2000, Vilone_2002}, they introduce additional and unnecessary computational complexity by requiring an extensive average over initial configurations, which exhibit a stochastically varying number of nominal categories. To overcome these issues and extend the model's realism, we generalize the Axelrod framework by introducing continuous cultural traits confined to the interval $[0,1]$. Crucially, the similarity threshold,  $d$ (which dictates that only traits within a distance $d$ are considered similar enough to interact), now serves as a continuous control parameter, enabling a robust analysis of the critical behavior using techniques like Finite-Size Scaling \cite{Privman_1990}. Moreover, this continuous formulation naturally maps the discrete problem into a geometric one, where $d$ is a direct measure of cultural tolerance, providing a transparent and quantifiable link between the local interaction rule and the resulting global phase behavior.

In this paper, we explore the dynamics of this continuous-trait Axelrod model in a two-dimensional square lattice with nearest-neighbor interactions under two distinct conditions: perfect copying and imperfect copying.  
Perfect copying refers to the standard Axelrod mechanism where an agent adopts the neighbor's exact trait value upon interaction. Imperfect copying introduces cultural drift, meaning the new trait is copied from a noisy distribution around the neighbor's value, reflecting mistakes or misinterpretations in transmission \cite{Mora_2025}.

For the perfect copying scenario, we precisely locate the non-equilibrium phase transition between the disordered,  fragmented phase ($d < d_c$) and the ordered polarized phase ($d > d_c$) at $d_c \approx 0.0784$.  These phases are characterized by the mean domain density ($\mu$) and the mean fraction of agents in the largest cultural domain ($\rho$).   A cultural domain (or cluster) is defined as a maximal connected component of agents who have become culturally compatible (i.e., the  traits of neighboring agents  are all within the similarity threshold $d$ of each other). In the thermodynamic limit,  the disordered phase is characterized by a finite domain density ($\mu>0$)  and vanishing largest domain size ($\rho=0$),  while the ordered phase  is characterized by a vanishing domain density ($\mu=0$) and  the coexistence of macroscopic domains ($0 < \rho <1$).
Crucially, we provide a definitive characterization of the critical behavior, which was previously unattainable in the discrete framework. 
Our analysis reveals a nuanced transition:  $\mu$   exhibits a continuous transition,  vanishing at $d_c$ with the  exponent $\beta =1/3$, while $\rho$ displays a discontinuous jump at $d_c$. Both behaviors are governed by the same exponent $\nu=2$, which describes the finite-size effects in the critical region.

For the imperfect copying scenario, we show that persistent noise fundamentally alters the dynamics by introducing a powerful selective pressure on the continuous trait space. This pressure leads to the emergence of two symmetry-related privileged attractors at  trait values $d$ and $1-d$.  However, we demonstrate that this noise-induced stabilization mechanism is extremely fragile: the attractors become unstable at large lattice sizes  or sufficiently large noise levels, which provides an explanation for the observed failure of the system to fully freeze into a consensus.  Our results collectively offer a complete picture of cultural dynamics in the continuous Axelrod model, from the precise characterization of its critical point to the identification of the selective forces governing its non-absorbing state under persistent cultural drift.

The remainder of the paper is organized as follows. In Section \ref{sec:mod}, we introduce our generalized Axelrod model with continuous cultural traits and define the simulation procedure.  Section \ref{sec:res} presents our results,   divided into two main subsections.  Subsection \ref{sub:1} analyzes the perfect copying scenario, focusing on the phase transition, the precise determination of the critical point ($d_c$),  and the Finite-Size Scaling analysis that yields the critical exponents $\beta$ and $\nu$.   This  analysis is  supported by  \ref{appA}, which provides the corresponding order parameters ($\mu$  and $\rho$)  for the discrete-trait Axelrod model.  Subsection \ref{sub:2} details the dynamics under imperfect copying, exploring the emergence and stability of the two privileged attractors.   Finally, Section \ref{sec:disc} summarizes our findings, discusses their implications for cultural evolution theory, and suggests avenues for future work.

\section{Model}\label{sec:mod}

In our continuous version of the Axelrod model, we assume that each agent $i=1, \ldots, L^2$  is characterized by F cultural traits $x_i^k \in [0,1]$ for $k=1,\ldots, F$.  Agents are fixed at the nodes of a square lattice of linear size $L$ with periodic boundary conditions and can interact with their four nearest neighbors only.

The core premise of the Axelrod model is homophily---the tendency for individuals to interact preferentially with similar others. In our continuous version, we define the similarity  $p_{ij}$  between two neighboring agents $i$ and $j$ as the fraction of their traits that are considered  `similar'. We assume that traits whose distance is less than or equal to a threshold $d$ are considered to be the same. The similarity is thus given by
\begin{equation}\label{pij}
p_{ij}  = \frac{1}{F} \sum_{k=1}^F \Theta \left (d - |x_i^k - x_j^k|  \right ),
\end{equation}
where $\Theta (z)$   is the Heaviside step function, defined as  $\Theta (z) = 1$ if $z \geq 0$ and $0$ otherwise.  An interaction between agents $i$ and $j$ occurs with a probability equal to their similarity, $p_{ij}$.

The simulation proceeds in discrete time steps. At each step, a focal agent, say agent $i$, is chosen at random. A neighbor, say agent $j$, is also chosen at random from its four nearest neighbors. An interaction then occurs with probability $p_{ij}$. If an interaction occurs, the focal agent $i$  attempts to copy one of the traits of its neighbor $j$.

The trait chosen for copying, say trait $k$, is selected at random from the subset of traits where the distance between the agent's and neighbor's traits ($ |x_i^k - x_j^k| $)  is greater than the threshold $d$. This means only traits that are not considered similar are subject to change.
The new trait of  the focal agent,  $\hat{x}_i^k $,  is a random variable distributed according to the truncated normal distribution  \cite{Mora_2025},
\begin{equation}\label{t_nor}
 f(\hat{x}_i^k|x_j^k)={\frac {1}{\sqrt{2\pi\sigma^2} }}\,{\frac {\exp (-{\frac {\hat{x}_i^k-x^k_j }{2\sigma^2 }})}{\Phi ({\frac {1-x_j^k }{\sigma }})-\Phi ({\frac {-x_j^k}{\sigma }})}},
\end{equation}
where  $ \Phi (z)={\frac {1}{2}}\left(1+\mbox{erf} (z/{\sqrt {2}})\right)$ is the standard normal cumulative probability function. Here the parameter $\sigma>0$  determines the accuracy of the copying process.   A smaller $\sigma$  results in the new trait being closer to the neighbor's trait, modeling the effect of social influence. This stochastic copying process is a key difference from the deterministic trait copying in the original model.

The model's behavior is driven by two key conditions related to the similarity $p_{ij} $:
\begin{itemize}
\item[-]  When $p_{ij} =0$, no traits are similar, so the agents do not interact and no cultural change occurs.
\item[-] When $p_{ij} =1$, all traits are similar (distance $\leq d$),  meaning there is no opportunity for cultural influence to take place. This corresponds to a state of complete cultural agreement between the two agents, and no further change can happen.
\end{itemize}
This dependence of interaction on similarity is the essential mechanism that leads to the emergence of distinct cultural domains, as observed in the original Axelrod model \cite{Axelrod_1997}. In that model, the competition between the disorder of the initial configuration that favors cultural fragmentation and the ordering bias of social influence that favors homogenization results in a nonequilibrium phase transition between those two classes of absorbing  configurations \cite{Castellano_2000,Vilone_2002,Reia_2016}.

To connect our continuous model to the original discrete one, we can relate their initial diversities. In the original model, where each trait can take on $q$  distinct values, the mean similarity of two randomly chosen agents is simply $1/q$.  For our continuous model, where traits are initialized from a uniform distribution on $[0,1]$, the probability that the distance between two traits is less than or equal to $d$ is $2d -d^2$. This is the mean similarity of a pair of agents in the initial configuration. By equating the initial mean similarities of both models, we can establish a correspondence between the number of discrete trait values $q$ and our continuous threshold parameter $d$:
\begin{equation}\label{map}
q = \frac{1}{d(2-d)} .
\end{equation}
This relationship provides a key insight into the model's behavior. When $d=0$, the mean similarity is $0$, which corresponds to $ q\to \infty$.  This implies that every agent is culturally unique, and no interactions can occur, leading to a fragmented, frozen state. Conversely, when $d=1$, the mean similarity is  $1$, which corresponds to $q=1$. This means all agents are initially identical, and the system is already in a single, homogeneous state.

Finally, a key consideration in our continuous-trait model is the definition of a cultural domain. In the original discrete-trait model, a cultural domain is typically defined as a bounded region in the lattice where all agents share the exact same culture. This definition is more complex in our continuous model, as non-neighboring agents within the same domain might have traits that differ by more than the threshold $d$.

To address this, we define a cultural domain as a connected component within an undirected contact graph \cite{Newman_2018}. This graph is constructed by placing a link between any pair of neighboring agents whose traits differ by less than the threshold $d$. This means that within a domain, every agent can be reached from every other agent by following a path of `culturally similar' neighbors.
The identification and characterization of these domains can be efficiently performed using the Hoshen-Kopelman algorithm \cite{Hoshen_1976}. This definition allows us to directly measure the cultural fragmentation of the system and compare our results to the phase transitions observed in the original Axelrod model.

\section{Results}\label{sec:res}

In our simulations, we analyze the system's behavior in both the perfect and imperfect copying regimes. In the case of perfect copying ($\sigma=0$), our copying rule is deterministic and guaranteed to lead the dynamics to an absorbing configuration.  
An absorbing configuration is a stationary state where all neighboring agents have reached a frozen state of cultural interaction, meaning that all their cultural trait differences are either all less than or all greater than the threshold $d$. In these configurations, no further cultural change can occur. Since no new traits can be created during the time evolution, the only possible trait values at any time are those present in the initial configuration.

In contrast, when copying is imperfect ($\sigma>0$), the stochastic nature of the copying rule introduces cultural drift,  which may prevent the system from reaching an absorbing configuration. The copying process, in which a new trait value is drawn from a truncated normal distribution centered on a neighbor's trait, acts as a continuous  source of cultural drift. This process continually produces novel cultural traits.  As a result, even large, culturally uniform domains become unstable and can erode over time due to this persistent drift, as demonstrated in the original discrete-trait model \cite{Klemm_2003}.  Consequently, the system may never freeze in an absorbing configuration, leading to sustained nonstationary dynamics.

When an absorbing configuration is reached, we quantify the system's cultural fragmentation by counting the total number of cultural domains  ($N$)  and measuring the number of agents in the largest domain  ($S_{max}$). These quantities are averaged over a large number of independent simulation runs (from $10^3$ to $10^5$ runs), which begin with different random initial cultural states (uniformly distributed on $[0,1]$) and distinct update sequences. For all results presented here, we focus on the case of two cultural traits, $F=2$.

Our main focus is on two macroscopic order parameters: the mean density of cultural domains
\begin{equation}\label{m}
\mu = \frac{ \langle N\rangle}{L^2}, 
\end{equation}
and the   mean fraction of agents in the largest domain
\begin{equation}\label{r}
\rho = \frac{ \langle S_{max} \rangle }{L^2}, 
\end{equation}
where $\langle \ldots \rangle$  denotes the average over independent simulation runs. These quantities are crucial because they indicate  how $\langle N\rangle$ and $\langle S_{max} \rangle$ scale  with the system size $L^2$  in the thermodynamic limit, thereby allowing us to characterize the different nature of the absorbing configurations.

\subsection{Perfect copying}\label{sub:1}

\begin{figure}[ht] 
\centering
 \includegraphics[width=1\columnwidth]{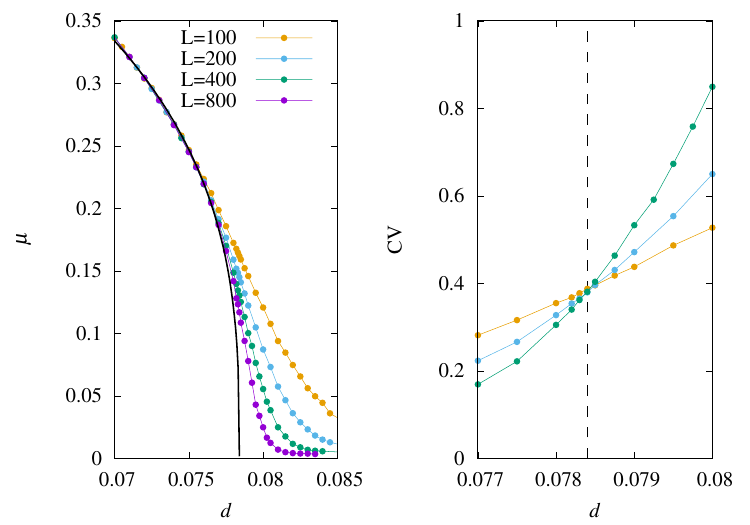}  
\caption{Mean density of domains $\mu$ (left panel) and coefficient of variation CV (right panel) as a function of the threshold $d$ for 
lattices of linear sizes $L=100, 200, 400$, and $800$ in the perfect copying case ($\sigma=0$).   The solid curve in the left panel  is a two-parameter fit function to the data for $L=800$, using the function 
$\mu = A(d_c-d)^\beta$.  The best-fit parameters are 
 $A= 1.643$ and $\beta = 0.335$,  with $d_c =0.0784$.
The dashed  vertical line in the right panel indicates the location  of  the critical point, $d_c \approx 0.0784$,   which is determined by the intersection of the CV data for the lattice sizes, $L=200$ and $L=400$. The error bars are smaller than the symbol sizes.}   
\label{fig:1}  
\end{figure}

In standard percolation theory, the density of clusters $\mu$  is a continuous and differentiable function of the model parameters and is  non-zero at the critical threshold \cite{Stauffer_1992}.  The situation is quite different in our model, as shown in the left panel of Fig. \ref{fig:1}.   Our results indicate  that for a  threshold $d$ greater than some critical value $d_c$,  the density of domains vanishes in the thermodynamic limit.  The vanishing density of domains, $\mu \to 0$,  implies the existence of only a few macroscopic domains in this region, which is a hallmark of an ordered (polarized) phase.  This  phase is characterized by  domains  of diverging mean size,  $1/\mu$. 

The transition at $d=d_c$  is continuous,  and the density of domains $\mu$ serves as an order parameter that vanishes at the critical point,  following the power law  $\mu \sim (d_c-d)^\beta$.     In continuous phase transitions, an efficient way to locate the critical point  $d_c$ is to analyze dimensionless ratios of moments of the random variable  $ N$,  such as the Binder cumulant \cite{Binder_2010}. Here,  we use  the coefficient of variation (CV) of the total number of domains, defined as 
\begin{equation}
\mbox{CV} = \sqrt{\frac{ \langle N^2 \rangle }{\langle N \rangle^2} -1},
\end{equation}
and  shown in the right panel of Fig. \ref{fig:1}.   The critical point  $d_c \approx 0.0784$ is identified as the intersection of the CV data for the  lattice sizes  $L=200$ and $L=400$, which is also very close  to the intersection of $L=100$ and $L=200$.  To obtain the smooth  data shown in this panel in the transition region,  we averaged over $10^5$ runs.   We note that data for $L=800$  is excluded from this plot because the inherent critical fluctuations---which are amplified by the CV measure, as it quantifies dispersion that diverges at $d_c$ --- made the location of the intersection point imprecise,  as averaging over such a large number of runs for this lattice size is computationally unfeasible.   The larger system size data for$L=800$will therefore be prioritized for analyzing the less noisy measures $\mu$ and $\rho$.  With $d_c$ determined, we perform a fit of the $\mu$  data for $L=800$ (left panel) to estimate the critical exponent $\beta$, which yields a value of $\beta = 0.335 \pm 0.005$.  This result is sufficiently close to the rational $1/3$, so we will assume $\beta = 1/3$ for the subsequent Finite-Size Scaling analysis.

\begin{figure}[ht] 
\centering
 \includegraphics[width=1\columnwidth]{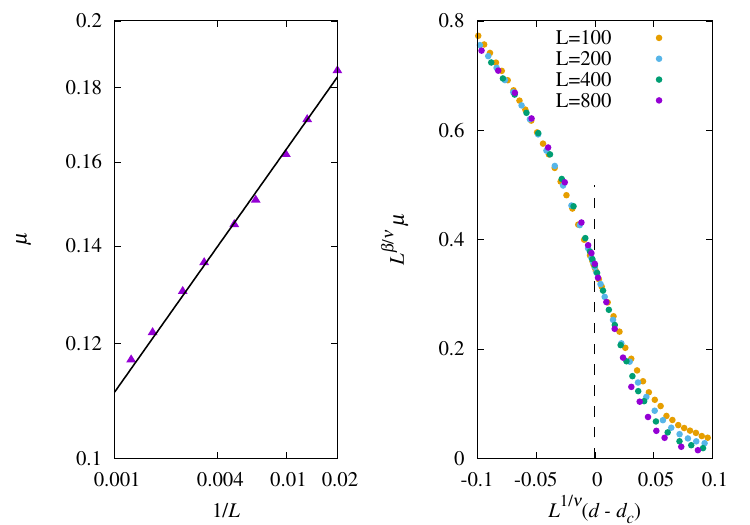}  
\caption{ (Left panel) Log-log plot of the mean density of domains $\mu$  at the critical threshold $d_c=0.0784$  as a function of  the reciprocal of the linear lattice size in the  perfect copying case  ($\sigma=0$). The curve fitting the data is $\mu =BL^{-\beta/\nu}$ with $B= 0.35 \pm 0.01$  and $\beta/ \nu = 0.167 \pm 0.006$. (Right panel) Scaled order parameter against the scaled distance to the critical threshold  for 
lattices of linear sizes $L=100, 200, 400$, and $800$. The parameters used are $\beta=1/3$, $\nu=2$, $d_c = 0.0784$, and $\sigma=0$.  
The error bars are smaller than the symbol
sizes.
}   
\label{fig:2}  
\end{figure}

The Finite-Size Scaling  theory asserts that for large $L$ and $d$ close to the critical threshold $d_c$,  the density of domains is governed by the scaling relation \cite{Privman_1990}:
\begin{equation}\label{m_L}
\mu = L^{-\beta/\nu} f \left [ L^{1/\nu}(d-d_c) \right ] ,
\end{equation}
where $\nu > 0$ is the  correlation length critical exponent and $f(z)$ is a scaling function such that 
$f(z) \sim z^\beta$ for $z \to - \infty$.   According to this scaling relation,  $\mu$  must decrease to zero as the power law  $\mu \sim L^{-\beta/\nu} $  at the critical point  $d=d_c$.  We explore this fact in the left panel of Fig.  \ref{fig:2}.  where $\mu$  is plotted
against $1/L$  on a log-log scale,  to determine the ratio $\beta/\nu$ and,  subsequently, the exponent $\nu$. We find a ratio of $\beta/ \nu = 0.167 \pm 0.006$,  which yields $\nu = 1.994 \pm 0.07$ using the previously determined value of $\beta$.  As before, this value is sufficiently close to $2$, so we will assume $\nu=2$. The validity of our conjectured exponents, $\beta =1/3$ and $\nu =2$,  is assessed by checking whether the scaled quantity $L^{\beta/\nu} \mu $ is independent of the lattice size $L$ when plotted against the scaled distance to the critical point $L^{1/\nu}(d_c-d) $,  as predicted by Eq.  (\ref{m_L}).  The quality of the collapse of the data for distinct lattice sizes is very good for $d < d_c$.  For $ d> d_c$,  the collapse is good only close to the critical threshold, which is expected since the scaling relation (\ref{m_L}) is strictly valid only within the critical region. 

\begin{figure}[ht] 
\centering
 \includegraphics[width=1\columnwidth]{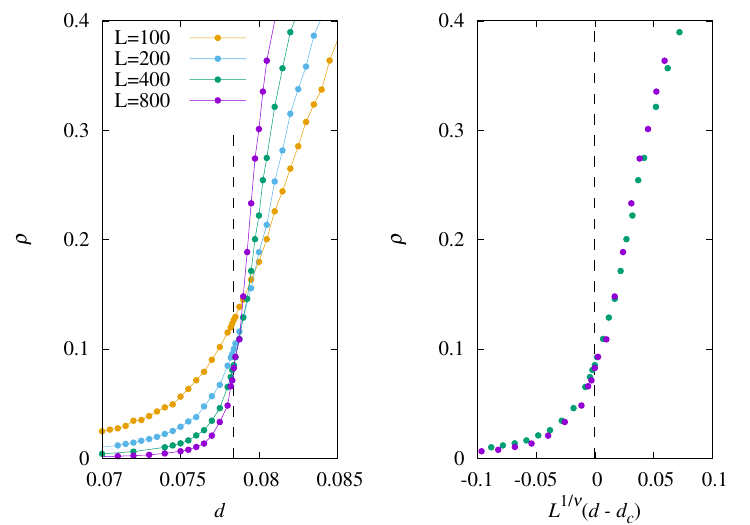}  
\caption{ (Left panel) Mean fraction of agents in the largest domain $\rho$  as a function of the threshold $d$ in the critical region for 
lattices of linear sizes $L=100, 200, 400$, and $800$ in the  perfect copying case  ($\sigma=0$).  (Right panel) $\rho$ against the scaled distance to the critical threshold  for 
lattices of linear sizes $L=400$ and $800$. The parameters used  for the data collapse are the critical exponent $\nu=2$ and the critical point $d_c = 0.0784$. The dashed  vertical lines indicate the location  of  the critical point $d_c$.
The error bars are smaller than the symbol
sizes.
}   
\label{fig:3}  
\end{figure}

The  mean fraction of agents in the largest domain $\rho $,  which is the usual order parameter in percolation problems \cite{Stauffer_1992},  exhibits a discontinuous phase transition at the critical threshold $d_c$ as shown  in the left panel of Fig.  \ref{fig:3}.   Specifically,  for $d  < d_d$ we have $\rho \to 0$ in the thermodynamic limit, indicating an extensive number of small domains with finite mean size  $1/\mu$.  At the critical threshold,  $\rho$ jumps abruptly from zero to a finite value $\rho_c < 1$ and then increases towards 1 with increasing $d$.  This jump is a definitive characteristic of a first-order (discontinuous) transition and marks the abrupt emergence of macroscopic cultural order through the nucleation of a few extensive domains.  The intersection of the curves for different lattice sizes confirms the transition location $d_c$. We note that while this crossing point serves to locate $d_c$ here, the crossing of the order parameter curves is not universally required for discontinuous transitions, as is the case in some models like the random geometric graph with periodic boundary conditions \cite{Dall_2002,Goel_2005}.

This  discontinuous  transition is surprising,  since $\rho$ exhibits a continuous transition from a disordered ($\rho=0$) to an ordered  ($\rho>0$) regime in the Poisson variant of Axelrod's model for $F=2$ \cite{Reia_2016}. However, 
the intersection point of the  $\rho$ curves for different $L$ varies significantly with $L$,  which is consistent with a discontinuous transition but makes it an unreliable method for precisely determining $d_c$.  In contrast, the crossing of the CV curves shown in Fig. \ref{fig:1} is robust, taking place at nearly the same value for all $L$.

 Assuming the scaling relation
\begin{equation}\label{r_L}
\rho = g \left [ L^{1/\nu} (d-d_c) \right ]
\end{equation}
 in the transition region,   the derivative of $\rho$ at $d=d_c$ is expected to diverge like $L^{1/\nu} $  as $L\to \infty$.  Thus,  the exponent $\nu$ determines the finite-size rounding (or sharpness) of the transition  for finite $L$ \cite{Kirkpatrick_1994, Campos_1998}.   We emphasize that this $\nu$  is a finite-size rounding exponent and is not the standard correlation length exponent, as the correlation length remains finite at a discontinuous transition.  While the numerical value of this  exponent could {\it a priori} be different from the one characterizing the continuous transition of $\mu$  (Eq. (\ref{m_L})),  we find them to be identical, and thus use the same notation $\nu$.   The scaling function $g(z)$ is such that $g(z) \to 0$  when $z \to - \infty$  and $g(z) \to \rho_c < 1$  when $z \to  \infty$.  
  Using  the  scaling relation (\ref{r_L})  for $\rho$ 
with the already determined parameters,  $d_c = 0.0784$ and $\nu =2$,  yields a very good collapse of the data for
\ $L=400$ and $L=800$,  as shown in the right panel of Fig. \ref{fig:3}.   In addition, we find $g(0) \approx 0.068$.   It is  highly reassuring that the exponent $\nu =2$ governs the finite size effects for both the continuous transition of $\mu$ and  the discontinuous transition of $\rho$, confirming the robustness of critical exponents as properties of the model  \cite{Stanley_1987}.

\begin{figure}[ht] 
\centering
 \includegraphics[width=1\columnwidth]{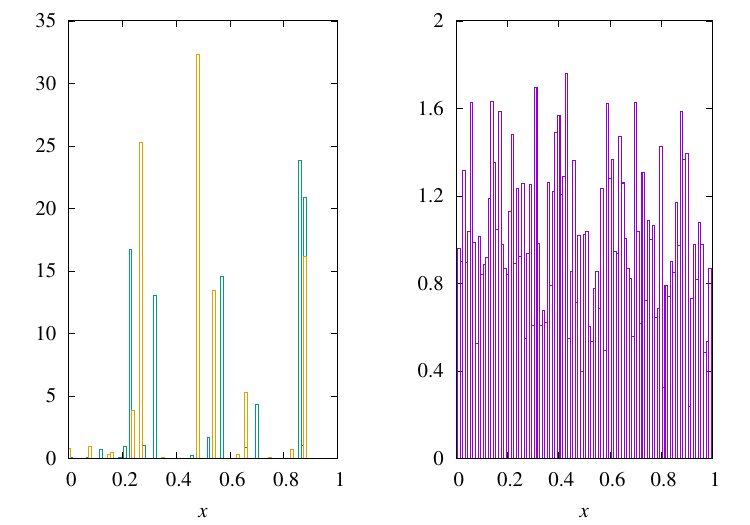}  
\caption{ Histogram of the cultural traits $x$ in the absorbing configurations for the perfect copying case ($\sigma=0$).   The results are shown for the ordered phase   for the two traits $k=1$ and $k=2$  in a single simulation run (left panel) and averaged over $100$  runs (right panel).  The parameters used are  $L=400$ and $d=0.1$. 
}   
\label{fig:4}  
\end{figure}

To conclude the analysis,  we consider the distribution of traits $x$ in  the absorbing configurations.  We recall that in the initial configuration the  traits are distributed uniformly in the unit interval.  Figure \ref{fig:4} shows the histograms of the final traits $k=1$ and $k=2$ in a single simulation and the average over $100$ runs  in the ordered phase ($d=0.1$).   Note that the values of the traits $k=1$ and $k=2$  evolve independently. 

 In any single run,  the initial uniformity of traits is broken, and the agents take on  a few distinct cultural values, as shown in the left panel of Fig. \ref{fig:4}.   This figure illustrates the microscopic basis for the coexistence of macroscopic domains.  For the parameters used in the figure, for instance, the largest domain contains on average only about  $57\%$ of the agents,  with the remainder distributed among other large domains. However, when we average the final trait distribution over many independent runs, the initial trait symmetry is restored, and the resulting histogram corresponds to a uniform distribution (right panel of Fig. \ref{fig:4}). As expected, the same uniform result holds when averaging over runs in the disordered phase ($d< d_c$).

\subsection{Imperfect copying}\label{sub:2}

Now we consider the case of imperfect copying, i.e. , $\sigma > 0$. In this scenario, the focal agent acquires a cultural trait that is similar to, but stochastically different from, their neighbor's trait. The parameter $\sigma$  of the truncated normal distribution (\ref{t_nor}) serves as a measure of the copying fidelity or inherent noise in the imitation process underlying social influence. While this scenario is arguably more realistic than perfect copying, the persistent noise introduces significant technical challenges, particularly concerning the system's convergence to an absorbing configuration, which we detail in this section.

\begin{figure}[t] 
\centering
 \includegraphics[width=1\columnwidth]{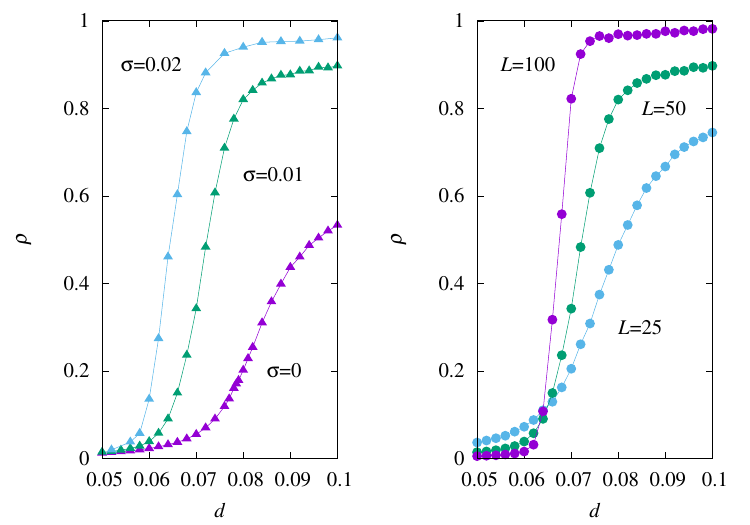}  
\caption{ Mean fraction of agents in the largest domain $\rho$  as a function of the threshold $d$.
The left panel shows   $\rho$ for a fixed lattice size ($L=50$) across different noise levels ($\sigma=0, 0.01$, and $0.02$).  The right panel shows $\rho$ for a fixed noise level ($\sigma=0.01$) across different lattice sizes ($L=25, 50$, and $100$). 
}   
\label{fig:5}  
\end{figure}

Figure \ref{fig:5} summarizes the results for the mean fraction of sites in the largest domain, $\rho $. The relevant finding is that the copying noise promotes the eventual dominance of a single cultural domain for $d$ greater than some critical threshold. This is indicated by the fact that $\rho $  approaches $1$  as $\sigma$ increases for a fixed lattice size $L$ (left panel) or as $ L$  increases for any fixed $\sigma > 0$ (right panel). This behavior contrasts sharply with the $\sigma =0$ case, where the ordered phase is characterized by the coexistence of multiple macroscopic domains, resulting in $ \rho < 1$ even in the thermodynamic limit. The introduction of noise thus leads to a more complete cultural homogenization.

The system's dynamics in the imperfect copying regime present a challenge for simulation time. The convergence to a frozen configuration in the disordered phase  ($d < d_c$) remains fast and reliable. However, in the ordered phase 
($d > d_c$), the freezing process is extremely slow and may not happen at all within practical simulation times.  This sluggish behavior is due to the persistent cultural noise ($\sigma > 0$) continually stirring the traits of the agents. 
To manage this, we impose a time limit of $2 L^2 \cdot 10^7  $   update attempts, discarding any run that does not freeze by this point. To optimize the simulation speed, we attempt to update only agents in active links (i.e., neighboring agents $i$ and $j$ for which the interaction probability is $p_{ij} > 0$ and $p_{ij} < 1$).
The rate of freezing failure increases with both the noise level $\sigma$  and the lattice size $L$.  This limitation explains why we could not produce data for higher noise levels (e.g., $\sigma=0.03$)  or larger system sizes (e.g.,
$L=200$)  in Fig. \ref{fig:5}.

\begin{figure}[t] 
\centering
 \includegraphics[width=1\columnwidth]{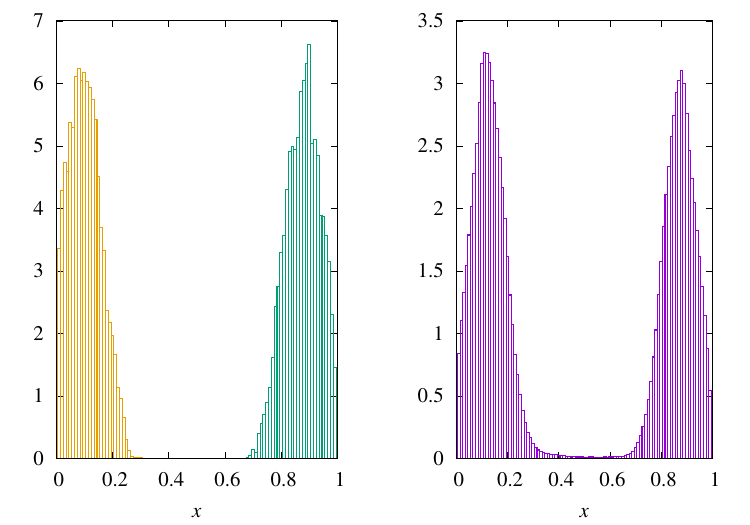}  
\caption{ Histogram of the cultural traits $x$ in the absorbing configurations for the imperfect copying case  ($\sigma=0.01$). The results for both traits ($k=1$ and $k=2$) are shown for the ordered phase ($d=0.1$): for a single simulation run (left panel) and averaged over $100$ independent runs (right panel). The lattice size is $L=100$ and the threshold is $d=0.1$.
}   
\label{fig:6}  
\end{figure}

The tendency of the dynamics to freeze into a single, homogeneous cultural domain in the ordered phase, as seen in Fig. \ref{fig:5}, can be understood by examining the distribution of cultural traits shown in Fig. \ref{fig:6}. Unlike the perfect copying scenario ($\sigma=0$), where the average trait distribution remains uniform, here the average over simulation runs yields a bimodal distribution (right panel of Fig. \ref{fig:6}). This indicates that the imperfect copying dynamics, driven by noise, exerts a selective pressure that privileges certain states. Specifically, the privileged states are those concentrated around $x=d$ and $x=1-d$. The left panel of Fig. \ref{fig:6} shows that in any single run, the dynamics break the initial uniformity, concentrating the traits around one of these  two values, which collectively form the bimodal distribution when averaged. 

We note that while the interaction probability depends on all traits, the copying mechanism for a selected trait $k$ is independent of the other traits. Since the concentration of traits is the result of stochastic symmetry breaking, a different simulation run could result in both traits $k=1$ and $k=2$ concentrating around the same privileged value (e.g., both around $x=d$). In such a case, the two histograms in the left panel of Fig. \ref{fig:6} would coincide.

This phenomenon can be quantitatively understood by calculating the probability that a noisy copy $\hat{x}_i^k=y$ is within a distance $d$ of the  original  trait $x_j^k = x$. This quantity, which we denote $P(x)$,  represents the probability of ``correctly''  copying the trait $k$ such that the new trait remains within the similarity threshold of the neighbor:
\begin{equation}
P(x) = \int_{\max(0,x-d)}^{\min(1,x+d)}  f(y|x)  dy, 
\end{equation}
where $f(y|x)$ is the truncated normal distribution,  Eq.  (\ref{t_nor}).  Explicit  evaluation of this integral yields  the following piecewise function,
\begin{equation} \label{P_piece}
P(x) = 
	\begin{cases}
	\left [ \Phi (d/\sigma) - \Phi (-x/\sigma)  \right ] /Z& \text{if $x <d$} \\
	 \left [ \Phi (d/\sigma) - \Phi (-d/\sigma)  \right ] /Z& \text{if $d<x <1-d$} \\
	\left [ \Phi ((1-x)/\sigma) - \Phi (-d/\sigma) \right ]/Z & \text{if $x >d$} ,
	\end{cases}
\end{equation}
where $Z= \Phi \left ( (1-x)/\sigma  \right )-\Phi \left  (-x/\sigma \right ) $ is the normalization factor from the truncated distribution,  and $ \Phi (z)$ is the standard normal cumulative probability function.  

\begin{figure}[t] 
\centering
 \includegraphics[width=1\columnwidth]{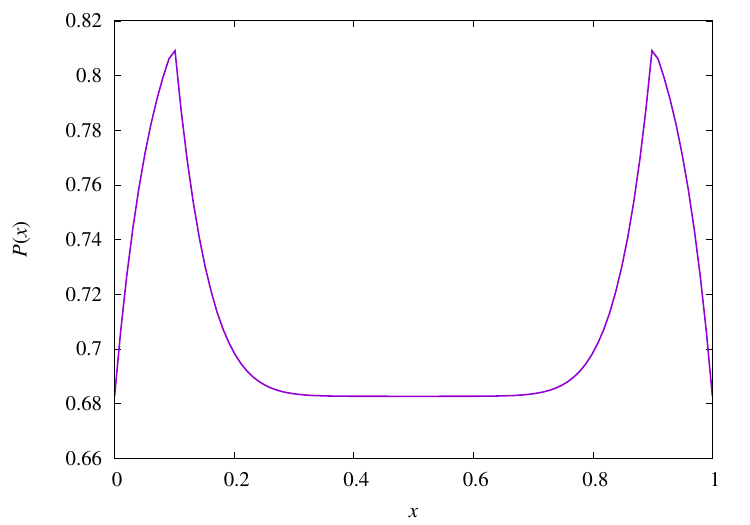}  
\caption{ Probability of copying correctly a trait of value $x$, $P(x)$,  for $d=0.1$ and $\sigma=0.1$.  The curve is generated from the theoretical piecewise function given by Eq. (\ref{P_piece}).
}   
\label{fig:7}  
\end{figure}

Figure \ref{fig:7} plots $P(x)$ for a relatively large noise parameter $\sigma=0.1$ to clearly emphasize the result that the probability of a ``successful'' copy, $P(x)$, is maximized at the  states $x=d$ and $x=1-d$. This advantage persists even as $\sigma$ is decreased (though the magnitude lessens). This selective pressure, favoring trait values that are maximally self-reproducible under the noisy copying mechanism, is the underlying reason why the dynamics consistently drive the system towards states concentrated around $d$ and $1-d$.

The finding that the ordered phase consists of two symmetry-related classes of absorbing configurations permits us to understand why the dynamics fail to freeze in some runs. Specifically, our analysis suggests that the freezing failure is due not to the competition between the domains with traits around $x=d$ and $x=1-d$, but rather to the weakness of the copying mechanism to maintain a cohesive group of agents against the constant influence of noise.

\begin{figure}[h] 
\centering
 \includegraphics[width=1\columnwidth]{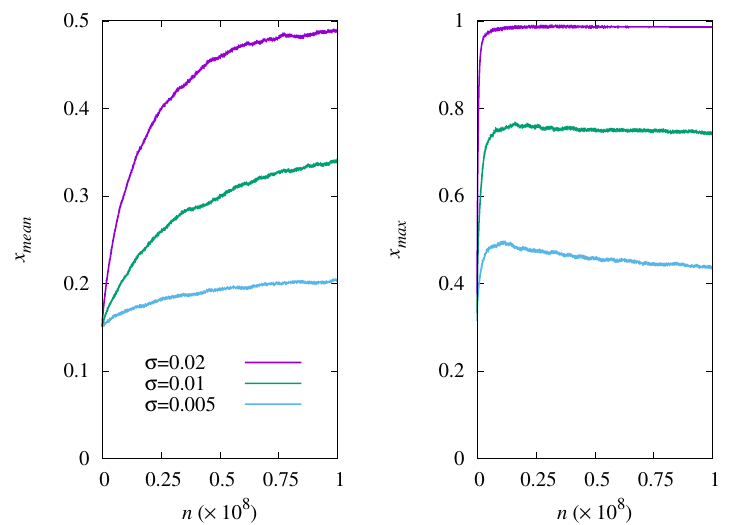}  
\caption{ The mean  trait value  $x_{mean} $  (left panel) and the maximum  trait value  $x_{max}$ (right panel) of the distribution of traits as a function of the number of active link update attempts $n$, for noise levels $\sigma=0.005, 0.01$, and $0.02$.  The results are averaged over $1000$ independent simulation runs, beginning with a uniform trait distribution in the isolated interval $(0,0.3)$.  The lattice size is $L=100$ and the threshold is $d=0.1$.}   
\label{fig:8}  
\end{figure}

To isolate this effect, we initialized our runs with all agents having traits concentrated close to one of the privileged states, $x=d$. For $d=0.1$, we chose the initial traits as uniform random numbers in the interval $(0,0.3)$. In this case, only about  $4/9 \approx 0.444$ of the links are active initially. Crucially, the maximum possible trait value is  $0.3$, ensuring that the initial configuration is far from the antagonistic configuration centered at $x=0.9$, thus eliminating competition at the initial stages of the dynamics. For each simulation run, we evaluate the mean trait value $x_{mean} = \sum_{i=1}^{L^2} x_i /L^2$ and the maximum trait value  $x_{max} =\max \{ x_i  \}$ as functions of the number of active link update attempts, $n$. We note that $n$ is not the number of discrete time steps introduced in Section \ref{sec:mod}: on average,  $2L^2/M$ discrete time steps are required to select an active agent, where $M$ is the current number of active links.

Figure \ref{fig:8} shows the evolution of  $x_{mean} $ (the mean trait value) and $x_{max}$ (the maximum trait value) averaged over $1000$ simulation runs.  
 It is clear that the stabilizing copying mechanism cannot maintain the traits near their initial values and, consequently, cannot stabilize the domain around the privileged state  $x=d$, unless the noise $\sigma$   is very small.  For instance, for  $\sigma =0.02$, the copying errors quickly propagate and generate new trait values that diffuse toward the antagonistic region centered at $x=1-d=0.9$.  This diffusion destabilizes the system and prevents any possibility of cultural freezing. 
The reason that  $x_{mean} $ does not tend toward $d=0.1$  even for the smallest noise value $\sigma=0.005$ (as the existence of the attractor might suggest from Fig. \ref{fig:6}) is due to a systematic bias originating from agents whose traits are already frozen in the initial configuration  (where $x_{mean} =0.15$) and remain so forever, thus biasing the final time-averaged mean trait to higher values.  This specific bias is not present when the initial traits are chosen uniformly across the unit interval.

\section{Discussion}\label{sec:disc}

In this study, we introduced and thoroughly  analyzed a generalized Axelrod model for the dissemination of culture that employs continuous cultural traits instead of the discrete, finite-state features used in the original formulation \cite{Axelrod_1997}.  This generalization aligns our framework with recent developments in cultural evolution theory, particularly the models of preference dynamics and intergenerational cultural transmission \cite{Bisin_2001,Buechel_2014,Mora_2025}.  Continuous traits---such as political opinions, beliefs, or lifestyle choice---are arguably more realistic representations of human culture,  as they typically exist on a spectrum (e.g., $[0,1]$) rather than being restricted to a few nominal categories (e.g., $\{ 1,2,3, \ldots \}$).   Furthermore, the continuous framework naturally allows the similarity threshold,  $d$, to be interpreted as a direct measure of cultural tolerance, providing a clearer quantitative link between local interaction rules and global outcomes.

A key advantage of this continuous-trait formulation emerges in the perfect copying scenario ($\sigma=0$). In this regime,   we were able to precisely locate and characterize the non-equilibrium phase transition from a disordered,  fragmented state ($d < d_c$) to an ordered,  polarized state  ($d > d_c$),  characterized by the coexistence of a few macroscopic cultural domains.  Our critical threshold finding,  $d_c \approx 0.0784$, corresponds to an effective critical state number of $q_c \approx 6.64$ using the mapping of Eq. (\ref{map}).  This result is reasonably close to the transition observed in the original discrete-trait model for $F=2$ (detailed in  the \ref{appA}), which transitions between an ordered phase for $q \leq 5$ and a disordered phase for $q \geq 6$.  Since the parameter $q$ is discrete, the original model is incapable of providing information on the nature of the phase transition or its critical exponents.

Our continuous-trait approach allowed for a robust analysis of the critical behavior of the model, providing a computationally efficient alternative to the Poisson variant of the original Axelrod model \cite{Castellano_2000, Vilone_2002, Reia_2016}. The Poisson variant is known to be computationally expensive due to the large number of samples required to average out the fluctuations produced by its stochastic initial conditions, particularly since the number of nominal categories changes with each initial condition.  Furthermore, for $F=2$, the Poisson variant yields a significantly different critical point ($q_c \approx 3.1$  \cite{Peres_2015}). 

Crucially, our analysis of the perfect copying scenario revealed a nuanced transition behavior:
\begin{itemize}
\item[-] The mean domain density,  $\mu$ (the order parameter), exhibits a continuous transition, vanishing at $d_c$
 with the exponent $\beta = 1/3$.
 \item[-] The fraction of agents in the  largest domain, $\rho$,  exhibits a surprising discontinuous transition. This contrasts with the Poisson variant, where $\rho$ transitions continuously \cite{Reia_2016}, suggesting that the noise inherent in the Poisson sampling process may smear out this underlying discontinuity.  
\end{itemize}
This combined behavior ($\mu \to 0$ continuously,  $\rho$ jumps) marks a hybrid change in the system's structural scaling: $d_c$ is a unique point where the fragmented phase collapses (as $\mu$ vanishes), and the macroscopic ordered phase emerges instantaneously (as $\rho$ jumps to a finite value $\rho_c$). This abrupt onset of long-range cultural ordering suggests a structural reorganization, rather than a classical first-order phase coexistence.
Interestingly, the finite-size effects for both the continuous transition of $\mu$  and the discontinuous transition of $\rho$  are consistently governed by the same exponent, $\nu=2$. This mirrors the phenomenology of hybrid phase transitions (discontinuity together with critical scaling) found in complex systems like interdependent-network percolation \cite{Lee_2016} as well as $k$-core percolation \cite{Baxter_2015}, albeit with the critical behavior unveiled here in domain statistics rather than avalanche observables.

In the imperfect copying scenario ($\sigma > 0$),  we find that the persistent noise prevents the system from reaching true absorbing configurations, a finding shared by other noisy Axelrod variants \cite{Klemm_2003, Perrier_2021}. More profoundly, we find that the noise exerts a selective pressure on the continuous trait space.  In any single run, the dynamics force the traits to   concentrate  around  one of two privileged states,  $x=d$  and $x=1-d$, though the overall population distribution (averaged over runs) is bimodal (Fig.  \ref{fig:6}).  Specifically, the states $x=d$  and $x=1-d$  are the trait values most likely to remain within the similarity threshold $d$ after a noisy copying event, acting as evolutionarily stable attractors that   promote cultural convergence toward one of the edges of the permissible cultural space. The final choice of the attractor is determined by stochastic  fluctuation biases set in the initial trait configuration. However, this stabilization mechanism is  very weak:  the attractors become unstable for large noise levels ($\sigma$) or even for smaller $\sigma$ if the lattice size ($L$) is sufficiently large. This instability leads to the observed failure of the dynamics to freeze, as the constant diffusive force of the noise overwhelms the weak stabilizing copying mechanism, preventing the trait distribution from being contained around the privileged states.

Finally, our model demonstrates that the generalization of the Axelrod framework to continuous cultural traits yields rich dynamics that directly address key open questions in cultural evolution. For the perfect copying case, the continuous formulation provides a robust and definitive characterization of the critical exponents ($\beta=1/3$, $\nu=2$), resolving ambiguities and inconsistencies present in discrete-trait model. More importantly, the inclusion of imperfect copying introduces a powerful mechanism of noise-driven selection, where the boundaries of the cultural space and the interaction rule conspire to create stable, privileged attractors at $x=d$ and $x=1-d$.  The ultimate failure of these attractors at large lattice sizes, where the dynamics cannot freeze, underscores the inherent fragility of cultural consensus when confronted with diffusive noise. This suggests that in large, continuously evolving societies, true cultural convergence to a monoculture is highly unlikely, giving way instead to sustained fragmentation or weakly confined domains. Future work could explore the effect of different noise distributions, alternative cultural spaces (e.g., circular traits), or the explicit incorporation of strategic imitation mechanisms, further bridging the gap between agent-based models and the theoretical framework of continuous cultural transmission \cite{Bisin_2001,Buechel_2014}.

\section*{Acknowledgments}

JFF is partially supported by  Conselho Nacional de Desenvolvimento Cient\'{\i}fico e Tecnol\'ogico  grant number 305620/2021-5.  PRAC was partially supported by Conselho Nacional de Desenvolvimento Cient\'{\i}fico e Tecnol\'ogico (CNPq) under Grant No. 301795/2022-3. This research is partially supported by Funda\c{c}\~ao de Amparo \`a Ci\^encia e Tecnologia do Estado de Pernambuco (FACEPE), Grant Number APQ-1129-1.05/24.



\appendix
\section{The discrete-trait Axelrod model for $F=2$}\label{appA}

\renewcommand{\thefigure}{A\arabic{figure}}
\setcounter{figure}{0}

\begin{figure}[ht] 
\centering
 \includegraphics[width=1\columnwidth]{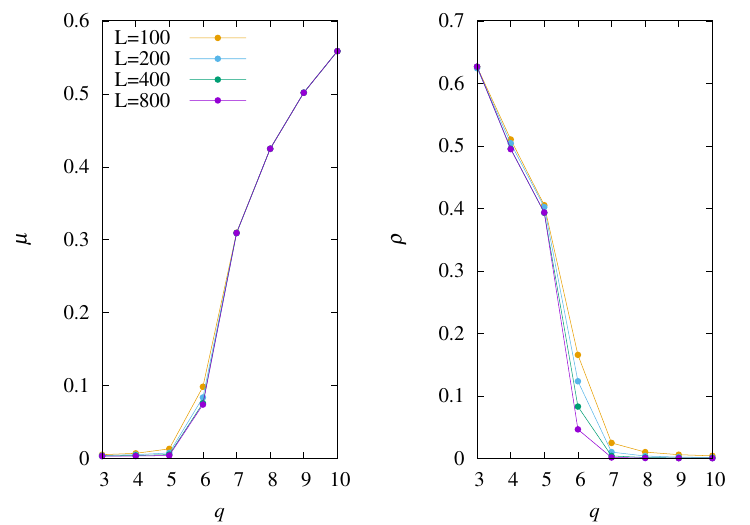}  
\caption{Mean density of domains $\mu$ (left panel) and  mean fraction of agents in the largest domain $\rho$  as a function of the  discrete number of traits $q$ for the original Axelrod model for 
lattices of linear sizes $L=100, 200, 400$, and $800$.  The results indicate the existence of an ordered phase for $q \leq 5$ and a disordered phase of $q\geq 6$. The error bars are smaller than the symbol sizes.}   
\label{fig:A1}  
\end{figure}

In this Appendix, we consider the original discrete-traits Axelrod model for $F=2$, in order to better contextualize our analysis of the critical region of our continuous-trait model in the case of perfect copying. Although the discrete-trait model, in which the states of the agents take on the integer values $x_i \in \{1,2,\ldots, q\}$, has been extensively considered in the literature (see, e.g., \cite{Klemm_2003b}), the analysis focused mainly on large values of $F$. To provide a direct comparison with our results, we carried out simulations of the discrete-trait model for $F=2$ and summarize the results in Fig. \ref{fig:A1}.

It is evident that, in the thermodynamic limit, a transition occurs between $q=5$ and $q=6$. For  $q \leq 5$, we observe an ordered phase where $\mu \to 0$ and $0 < \rho <  1$, corresponding to the coexistence of large domains of distinct cultures---the polarized state famously discussed by Axelrod. Conversely, for $q \geq 6$, we find a disordered, fragmented phase, where infinitely many cultural diverse microscopic domains coexist ($\mu > 0$ and $\rho \to 0$).

Due to the discreteness of the parameter $q$, we cannot rigorously determine the exact nature or critical exponents of the transitions in the original model. Nevertheless, the results suggest a non-trivial scenario. The mean domain density $\mu$ appears to undergo a transition that is relatively smooth, given that the change from a small finite value at $q=6$ to $\mu =0$  at $q=5$ is modest. In contrast, the mean fraction of agents in the largest domain $\rho$  appears to undergo a clear discontinuous transition, as $\rho$ jumps sharply from $\rho \approx 0.4$ at $q=5$  to $\rho \to 0$  at $q=6$.  This coexistence of a continuous-like change for $\mu$  and a discontinuous jump for $\rho$ is remarkably consistent with our findings for the continuous-trait model in the perfect copying regime.  We note that the absence of curve crossing for $\rho$ with increasing lattice size is consistent with discontinuous transitions observed in other systems, such as the random geometric graph with periodic boundary conditions \cite{Dall_2002,Goel_2005}.

Another interesting result arises from comparing the two models using the mapping in Eq. (\ref{map}). When $d$ in the continuous-trait model is chosen such that the mapping yields an integer value of $q$, we find that for all $q$, the mean domain density $\mu$  of the continuous-trait model is lower than that of the discrete-trait model, and the mean fraction of agents in the largest cluster $\rho$ is greater. This implies that, for the same effective $q$, cultural domains are consistently larger in the continuous-trait model. This is not unexpected: the continuous nature of the trait space provides a denser set of intermediate cultural values, which facilitates the creation of connecting paths between agents and the spread of cultural influence, ultimately leading to enhanced cultural convergence and larger domains.

\end{document}